# Probing the Electron States and Metal-Insulator Transition Mechanisms in Atomically Thin MoS$_2$ Based on Vertical Heterostructures

Xiaolong Chen[1], Zefei Wu[1], Shuigang Xu[1], Lin Wang[2], Rui Huang[1,3], Yu Han[1], Weiguang Ye[1], Wei Xiong[1], Tianyi Han[1], Gen Long[1], Yang Wang[1], Yuheng He[1], Yuan Cai[1], Ping Sheng[1], Ning Wang[1]*

[1]*Department of Physics and the William Mong Institute of Nano Science and Technology, the Hong Kong University of Science and Technology, Clear Water Bay, Hong Kong, China*

[2]*Department of Condensed Matter Physics, Group of Applied Physics, University of Geneva, 24 Quai Ernest Ansermet, CH1211 Geneva, Switzerland*

[3]*Department of Physics and Electronic Engineering, Hanshan Normal University, Chaozhou, Guangdong, 521041, China*

*Correspondence should be addressed to: Ning Wang, phwang@ust.hk

**The metal-insulator transition (MIT) is one of the remarkable electrical transport properties of atomically thin molybdenum disulphide (MoS$_2$). Although the theory of electron-electron interactions has been used in modeling the MIT phenomena in MoS$_2$, the**

**underlying mechanism and detailed MIT process still remain largely unexplored. Here, we demonstrate that the vertical metal-insulator-semiconductor (MIS) heterostructures built from atomically thin $MoS_2$ (monolayers and multilayers) are ideal capacitor structures for probing the electron states in $MoS_2$. The vertical configuration of MIS heterostructures offers the added advantage of eliminating the influence of large impedance at the band tails and allows the observation of fully excited electron states near the surface of $MoS_2$ over a wide excitation frequency (100 Hz-1 MHz) and temperature range (2 K- 300 K). By combining capacitance and transport measurements, we have observed a percolation-type MIT, driven by density inhomogeneities of electron states, in the vertical heterostructures built from monolayer and multilayer $MoS_2$. In addition, the valence band of thin $MoS_2$ layers and their intrinsic properties such as thickness-dependence screening abilities and band gap widths can be easily accessed and precisely determined through the vertical heterostructures.**

Molybdenum disulphide ($MoS_2$), a n-type semiconductor[1-16], shows novel properties such as superconductivity[6], controllable valley polarization[17, 18] and metal-insulator transition[3-6] (MIT). In $MoS_2$ field-effect transistors (FETs), gate-induced charge carriers transport in a thin layer near the surface of $MoS_2$ and are vulnerable to charge impurities and different types of disorder[2, 3, 12, 14, 19, 20]. The presence of a high-k dielectric material[3] to monolayer $MoS_2$ can effectively screen charge impurities and allow the observation of MIT. Based on recent transport measurements[3], the phase transition behavior of monolayer $MoS_2$ has been attributed to transition from an insulating phase, in which disorder suppresses the electronic interactions, to a metallic phase in which strong coulomb interactions occur. However, the underlying physical mechanism and detailed MIT process need to be further clarified. Different from the studies on transport properties of MoS$_2$, the capacitance spectroscopy [14] recently applied to the characterization of MoS$_2$ FET structures has been demonstrated as one of the most convenient and powerful method for studying the electron states in MoS$_2$ at room temperature. At low temperatures, however, the information obtained by this technique is limited due to the large impedance near the band edge of MoS$_2$. Different from that in graphene quantum capacitors[21-27], the slow charge carrier mobility in MoS$_2$ capacitors often leads to incompletely charged states, mainly due to the localization near the band edge. The incompletely charged capacitance confuses the effect of charge traps.

Here, we show an approach to address these problems by introducing a MoS$_2$-based vertical metal-insulator-semiconductor-metal (MIS-M) heterostructure suitable for probing electron states using capacitance measurements. Unlike conventional FET structures[14], our approach eliminates the impedance effects and can directly access the intrinsic characteristics of thin-layer MoS$_2$ over a wide frequency (100 Hz- 1 MHz) and temperature range (2 K-300 K). By combing capacitance and transport measurements, we show that the MIT observed in monolayer and multilayer MoS$_2$ is consistent with the physical picture of a percolation[28-35] transition model. The results of our investigation on the mechanisms of MIT and other intrinsic characteristics, such as thickness-dependent screening abilities and fast relaxation of hole carriers at the valence band, provide useful information much needed for improving the performance of the FET devices based on MoS$_2$ monolayers and multilayers.

**MoS$_2$ vertical heterostructural capacitance devices**

Figs. 1a and 1b illustrate our specially designed MIS-M capacitor device, fabricated by transferring[23, 36] exfoliated flakes of MoS$_2$ and hexagonal boron nitride (BN) on a Si substrate coated with a SiO$_2$ thin layer (300 nm). Exfoliated natural crystals of monolayer or multilayer MoS$_2$ were first transferred onto a BN sheet, serving as an ultra-smooth and disorder-free gate dielectric[37]. A Ti/Au (10 nm/20 nm) local gate sits underneath the BN sheet. The critical step in achieving MIS-M structure is to have the MoS$_2$ sheet fully covered by a top electrode (Ti/Au: 10 nm/50 nm). The equivalent circuit of this device geometry is shown in Fig. 1c. The measured capacitance $C_t$ is the total capacitance contributed by two capacitors originating from the BN layer ($C_{BN}$) and MoS$_2$ ($C_{MoS}$) in serial connection, plus the residue capacitance $C_p$ in parallel connection. $C_t$, shown below, is the capacitance wiping off $C_p$ (see detailed analysis in Supplementary Material). Therefore, $C_t = (C_{MoS}^{-1} + C_{BN}^{-1})^{-1}$. With fully-covered top electrodes, carriers can respond vertically instead of moving in the plane of MoS$_2$. This unique structure directly avoids the huge lateral resistance $R$ of MoS$_2$ near the band edge. As a result, the measured $C_t$ (of a 5.9 nm-thick MoS$_2$) at 2 K (Fig. 1d) is almost independent of excitation frequencies $f$, which differs greatly from that obtained in conventional FET structure devices[7,14]. In capacitance measurement of conventional FET structures, lateral resistance $R$ must be considered when $R \sim 1/(2\pi f C_t)$. This is confirmed by our MoS$_2$ capacitance devices with partially-covered top electrodes which show significant frequency-dependent and temperature-dependent characteristics (Supplementary Material). We also achieved good Ohmic contacts between the top Ti/Au electrode and MoS$_2$ in our devices, as evidenced by the capacitance measurements at different excitation voltages (Fig. 1e). Note that the capacitance measured at large excitation voltages (e.g. at 2V) shows deviation due to the averaging effect.

**Characterization of the vertical MIS-M structures**

The interface structure and band diagrams in the MoS$_2$-based MIS-M devices are shown schematically in Figs. 4a-d. When the gate voltage $V_g > 0$, electrons accumulate at the MoS$_2$ surface (Fig. 4b). The measured capacitance approaches $C_{max} = C_{BN} = \frac{\varepsilon_{BN}}{d_{BN}}$ when $V_g$ is sufficiently large, where $\varepsilon_{BN}$ and $d_{BN}$ are the dielectric constant and the thickness of BN respectively, while under negative $V_g$ (Fig. 4c) electrons are depleted. In this case, the measured capacitance can be described by $C_{min} = (\frac{d_{BN}}{\varepsilon_{BN}} + \frac{d_{MoS}}{\varepsilon_{MoS}})^{-1}$, where $\varepsilon_{MoS}$ and $d_{MoS}$ are the dielectric constant and the thickness of MoS$_2$ respectively. This allows us to directly obtain the $\varepsilon_{MoS}$ - $d_{MoS}$ relationship of MoS$_2$.

As shown in Fig. 2, $\varepsilon_{MoS}$ has been found to increase from 3.9$\varepsilon_0$ for a monolayer to 10.5$\varepsilon_0$ for bulk MoS$_2$. This is in excellent agreement with theoretical predictions[38, 39]. The small $\varepsilon_{MoS}$ in monolayer MoS$_2$ suggests poor dielectric screening of Coulomb interactions, indicating that strong electron-electron interactions could be achieved in clean monolayer MoS$_2$. The largely increased mobility observed in monolayer MoS$_2$ placed in a high-$\kappa$ dielectric environment[2, 3, 40] probably benefits from its small $\varepsilon_{MoS}$. In fact, the decrease in the optical phonon mode $E_{2g}^1$ observed by a Raman spectroscopy study of few-layer and bulk MoS$_2$[41] is also due to the strong dielectric screening effects.

The valence band of multilayer MoS$_2$ is also accessed by detecting the inversion layer of holes using low excitation frequencies at sufficiently high temperatures (Fig. 3a and b). However, the inversion layer is invisible when using high frequencies at low temperatures (T<100 K). This is due to the presence of the Schottky-barrier between the Ti/Au contact and the valence band[7].

Holes must form through thermal excitations or minute current leakage into the contacts. This process often requires a long time from ~ms to seconds. In the 12 nm-thick MoS$_2$ capacitance device, the majority of hole carriers have been relaxed around 20 kHz at 300 K, as confirmed by the phase information of the device, which is defined by $\Theta = \arctan(G/2\pi f C_t V)$, where $G$ is the conductance and $V$ is the excitation voltage. As shown in Fig. 3c, the phase peaks appear at about 20 kHz for different negative gate bias voltages, indicating that the relaxation time of holes in the 12 nm-thick MoS$_2$ device is around 50 μs. For the capacitance samples contacted by Cr/Au top electrodes (Cr has a larger work function (~4.5 eV) than that of Ti (~4.3 eV)), a short relaxation time (~5 μs) for holes has also been achieved at 300 K (Supplementary Material).

By applying the Poisson equation to model the vertical heterostructures in a quasi-quantitative manner (Supplementary Material), we correlated the quantum capacitance $C_q$ of MoS$_2$ with the surface potential $V_s$ in our capacitance devices. $V_s$ is extracted based on the charge conservation relation $V_s = \int_0^{V_g} (1 - \frac{C_t}{C_{BN}}) dV_g$. The $C_q$-$V_s$ relation is shown in Fig. 3d, yielding a band gap around 1.14 eV, which is close to the reported value of 1.2 eV[42].

**Percolation-induced MIT in monolayer and multilayer MoS$_2$**

Similar to the MIT observed in transport measurements[3-6], the capacitance data of the 5.9 nm-thick MoS$_2$ device measured at different temperatures (Fig. 4m) show an interesting transition with a well-defined cross-over point (at $V_g$=5V and corresponding to a carrier density $n$~$6.8 \times 10^{12}$ cm$^{-2}$, obtained from $n = C_{BN}(V_g - V_s - V_T)/e$, where $V_T$ ~ −1V is the threshold voltage). When $V_g < 5$V, $C_t$ decreases with decreasing temperature, whereas at $V_g > 5$V the temperature dependence of $C_t$ is reversed. The observed cross-over point in capacitance measurements is indeed related to the MIT as its value ($\sim 6.8\times10^{12}\,\text{cm}^{-2}$) is consistent with that measured by transport (Ref. 6 and Fig. 6). More evidence is provided by capacitance measurements of monolayer $MoS_2$ samples (Fig. 5a). In monolayer $MoS_2$, the intersections of the capacitance curves showed obvious temperature-dependent characteristics. At temperatures below 100 K, we observed that the cross-over point was stabilized roughly at $n\sim1.2\times10^{13}\,\text{cm}^{-2}$, consistent with the transport results measured in monolayer $MoS_2$[3] with a MIT at $n\sim1\times10^{13}\,\text{cm}^{-2}$. Bilayer and trilayer $MoS_2$ samples displayed similar transition phenomena with cross-over points around $n\sim8.6\times10^{12}\,\text{cm}^{-2}$ (Fig. 5b).

The electronic transport of $MoS_2$ suffers from charge impurities[2,3] and short-range disorders[12,14,19,20], such as ripples, dislocation and sulphur vacancies. These disorders result in the insulating transport behaviour of $MoS_2$ in the low carrier density region, where electrons transport through hopping between localized states (Fig. 4j) and can be well described by the variable-range-hopping model[3,12,20]. In the region where sufficient large carrier densities are introduced, metal behaviour is observed[3]. Here, we propose a percolation-type MIT in $MoS_2$, driven by density inhomogeneity of electron states[28-35] which describes the systems in which charge carriers are transported through percolating conductive channels in the disorder landscapes due to the poor screening effect at low carrier densities. When carrier density is low enough, conductive paths are efficiently blocked and MIT occurs. $MoS_2$ has been proven to be such a disordered system, with impurity concentration ranging from $10^{11}\,\text{cm}^{-2}$ to $10^{13}\,\text{cm}^{-2}$, especially for monolayer $MoS_2$, which is more vulnerable to ripples and charge impurities[2,3,12,14,19,20]. Thus, the MIT in $MoS_2$ is in line with the percolation transition theory in which disorder plays an important role. Moreover, our capacitance and transport data, shown below, provide further evidences to this effect.

The evolution of concentration and effective thickness of electron states probed by capacitance measurements can explain the observed MIT in transport measurements fairly well and provide details of the percolation transition process. The percolation transition phenomenon is illustrated in Fig. 4j-k. With increasing carrier densities $n$ (by increasing gate voltage), the localized electron states begin to percolate with each other till a conductive channel occurs at a critical density (Fig. 4k). Further increasing carrier densities will lead to sufficient conductive channels spanning the entire system and result in metal-like transport behaviours (Fig. 4l). On the other hand, at the same carrier density, the effective thickness $d_{eff} = \frac{\varepsilon_{MoS}}{C_{MoS}}$ of electron states confined in the surface of MoS$_2$ can be tuned by varying temperatures (Fig. 4n). Smaller $d_{eff}$ can also be achieved at higher gate voltages where large amounts of surface charges are induced (supported by theoretical calculations in the Supplementary Material). As illustrated in Fig. 4e-i (assuming $n$ remains unchanged), more conductive channels are formed at a smaller $d_{eff}$. The MIT should occur when $d_{eff}$ is sufficiently small. The capacitance data of our samples (Fig. 4n and o) are similar to those obtained from transport measurements in multilayer MoS$_2$ (showing a MIT at $n=6.7\times10^{12}\,\text{cm}^{-2}$)[6]. When $n < 6.8\times10^{12}\,\text{cm}^{-2}$, $d_{eff}$ decreases with increasing temperatures. Hence, the conductivity should increase as the temperature increases. In contrast, when $n > 6.8\times10^{12}\,\text{cm}^{-2}$, the increase of $d_{eff}$ would lead to decreasing conductivity as the temperature increases. Furthermore, the increasing $n$ and decreasing $d_{eff}$ would also enhance the screening of disorders and electron states and thus lead to increasing conductivity, while lowering the Coulomb interaction strength.

The percolation transition also suggests an increasing transition density at the cross-over point with increasing impurity concentration[30-32]. In our MoS$_2$ samples the transition density was in the range $10^{12}$-$10^{13}$cm$^{-2}$ due to the presence of large amounts of impurities. Moreover, the transition density in monolayer MoS$_2$ ($\sim 1\times 10^{13}$cm$^{-2}$) was larger than that observed in multilayer MoS$_2$ ($\sim 6\times 10^{12}$cm$^{-2}$), which agreed with the prediction of percolation theory as monolayer MoS$_2$ was more vulnerable to disorders. This was further evidenced by extracting charge trap densities, $D_{it}$, of MoS$_2$ from capacitance measurements (a trilayer sample shown in Fig. 5c). The presence of impurities or disorder may cause charge-trapping effects in MoS$_2$ capacitance devices, particularly at low temperatures. The charge traps can be fully excited only at relatively low frequencies (e.g. 100 Hz). The density of the charge traps can then be estimated by measuring the difference in capacitance at low and high frequencies, i.e., $D_{it} = (C_{MoS}(low\_f) - C_{MoS}(high\_f))/e$. The trap densities in our monolayer and trilayer MoS$_2$ samples were in the order of $10^{12}$eV$^{-1}$cm$^{-2}$ (Fig. 5d). The trap densities in monolayer MoS$_2$ were apparently large, suggesting that monolayer MoS$_2$ is more sensitive to disorder. In fact, the trap densities in our samples were underestimated because of the limitation of the excitation frequency ranges. At relatively high temperatures (inset of Fig. 5c), the charge traps were easily excited, and the capacitances measured at low and high frequency show no difference.

At the transition point $n\sim 1.2\times 10^{13}$cm$^{-2}$ in monolayer MoS$_2$, a large ratio $r_s = \dfrac{m^* e^2}{2\pi\varepsilon_{MoS}\hbar^2 \sqrt{\pi n}} \approx 7.1$ of Coulomb energy and kinetic energy was obtained, with measured $\varepsilon_{MoS} = 3.9\varepsilon_0$ and effective electron mass $m^* = 0.45 m_0$ [3, 11]. However, in such strong interacting systems, negative compressibility was not observed by capacitance measurements. This is probably due to the suppression of Coulomb interactions by large amounts of impurities present in MoS$_2$. Thus the density inhomogeneity induced by the impurities should still dominate the properties at the transition point.

The percolation induced MIT in MoS$_2$ is further supported by transport data at low temperatures. The MITs of multilayer (Fig. 6a) and monolayer (Fig. 6b) MoS$_2$ are clearly shown by the conductivity $\sigma$, at different temperatures, similar to previous reports[3-6]. The MIT occurs at $n \sim 6\times10^{12}$ cm$^{-2}$ for multilayer MoS$_2$ and $n \sim 1.1\times10^{12}$ cm$^{-2}$ for monolayer MoS$_2$, consistent with the capacitance data. To gain further insight into the transition behavior, we applied the percolation model of conductivity[31, 32, 35] near the percolation threshold density $n_c$, which is described by

$$\sigma = A(n-n_c)^\delta$$

where $A$ is a constant of proportionality and $\delta$ is the percolation exponent. Below the threshold density $n_c$, the 2D electron gas broke up into isolated puddles of carriers with no conducting channels crossing the whole sample. The conductivity showed insulating behavior and eventually vanished at $T = 0$ K. In 2D systems, $\delta$ is expected to be 4/3[30, 31]. Based on the percolation model, we fit our experimental data of multilayer (Fig. 6c) and monolayer (Fig. 6d) MoS$_2$ samples at 2 K. The experimental results show excellent agreement with theoretical predictions. The extracted parameters are $\delta \sim 1.7$ and $n_c \sim 3.2 \times 10^{12}$ cm$^{-2}$ in multilayer MoS$_2$ and $\delta \sim 1.8$ and $n_c \sim 3.8 \times 10^{12}$ cm$^{-2}$ in monolayer MoS$_2$. The obtained percolation exponents are consistent with experimental values $\delta = 1.4$-$1.7$ found in other 2D systems, such as GaAs/AlGaAs heterostructures[31, 32]. There was a slight deviation between experimental data and fitting curves at low carrier densities due to enhanced hopping conductivity and quantum tunneling at finite temperatures.

**Conclusions**

The vertical MIS heterostructures built from atomically thin MoS$_2$ are ideal capacitor structures for probing the electron states and intrinsic properties of MoS$_2$. Based on the analyses of experimental data obtained by electrical transport measurement and capacitance spectroscopy, we believe that the percolation-type MIT (driven by density inhomogeneities of electron states) is the dominating mechanism of the MIT in both monolayer and multilayer MoS$_2$. The vertical heterostructures offer the added advantages of eliminating the influence of large impedance at the band tails and accessing intrinsic characteristics such as thickness-dependence dielectric constant and band gap variation in atomically thin MoS$_2$. The present study also provides a new approach to characterizing the intrinsic properties of other atomically thin-layered materials and interface states of heterostructures built from 2D materials.

**Methods**

**Sample preparation.** Monolayer and multilayer MoS$_2$ flakes were exfoliated from molybdenum disulphide crystals (from 2D Semiconductors) by the micromechanical cleavage technique. MoS$_2$ and BN flakes were placed on the surface of a glass slide coated with PDMS/MMA as described for graphene-BN device fabrication[36]. Then, these thin flakes were transferred onto a local Ti/Au (10nm/20nm) gate. The top electrodes were patterned using standard electron-beam lithography. Two types of top electrodes, Ti/Au (10nm/50nm) and Cr/Au (2nm/50nm), were fabricated through electron-beam evaporation. The dielectric constant of the BN sheet is measured by calibrating an internal reference capacitor that sits near the MoS$_2$ capacitance device (Fig. 1a). The thicknesses of MoS$_2$ and BN flakes were measured by an atomic force microscope (Veeco-Innova).

**Capacitance and transport measurements.** Capacitance measurements were carried out using an HP Precision 4284A LCR Meter with a sensitivity of ~0.1 fF in a cryogenic system (2 K- 300 K). All wires in the measurement circuits were shielded and the p-Si substrates were also grounded to minimize residual capacitance. The residual capacitance in the measurement setup is at the order of 1 fF (see Supplementary Material). Transport measurements were performed in the same cryogenic system using lock-in techniques.

## Acknowledgements

The authors are grateful for fruitful discussions with Prof. Z. Q. Zhang from HKUST. Financial support from the Research Grants Council of Hong Kong (Project Nos. HKU9/CRF/13G, 604112, HKUST9/CRF/08 and N_HKUST613/12) and technical support of the Raith-HKUST Nanotechnology Laboratory for the electron-beam lithography facility at MCPF (Project No. SEG_HKUST08) are hereby acknowledged.

## Author contributions

X. C. is the main contributor who initiated and conducted most experiments including sample fabrication, data collection and analyses. N. W. is the principle investigator and coordinator of this project. X. C., N. W., and P. S. provided the physical interpretation and wrote the manuscript. Other authors provided technical assistance in sample preparation, data collection/analyses and experimental setup.

## Competing financial interests

The authors declare no competing financial interests.

**Figures:**

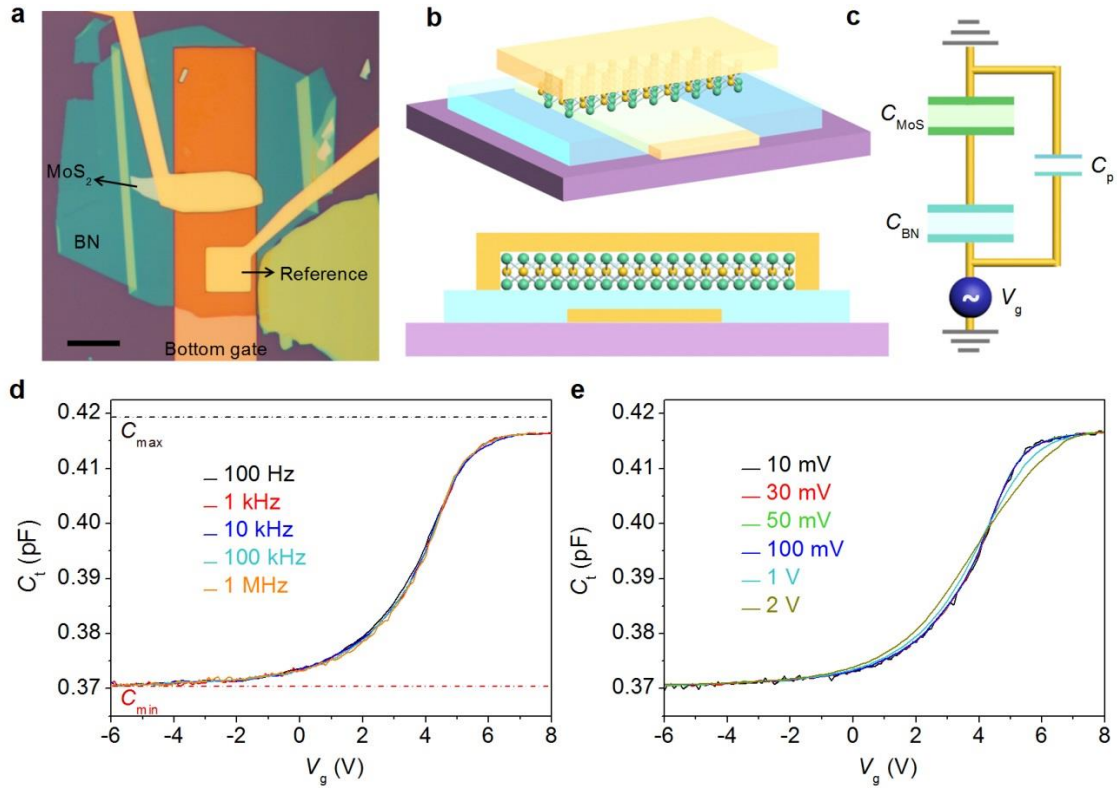

**Figure 1 | The optical and schematic images of the MoS₂ MIS-M heterostructures. a,b,** The MoS$_2$ flakes are fully covered by a top Ti/Au electrode. The square top electrode in (**a**) is the reference capacitor. Scale bar is 10μm. **c**, The equivalent circuit of the MoS$_2$ capacitance devices. **d,e**, Total capacitance $C_t$ measured from a 5.9 nm-thick MoS$_2$ at 2 K at different frequencies (**d**) and excitation voltages (**e**) respectively. The measured capacitance in vertical heterostructures is almost independent of excitation frequencies, which differs greatly from that obtained in conventional FET structures (Supplementary Material). The excitation voltage used for (**d**) is 50 mV, and the frequency used for (**e**) is 100 kHz.

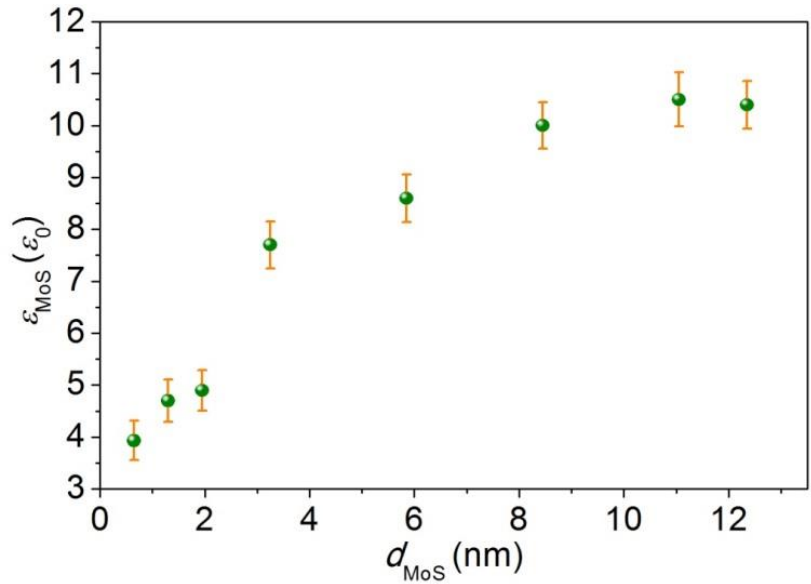

**Figure 2 | Experimental data of the thickness-dependent dielectric constant of MoS₂.** The dielectric constant of MoS₂ $\varepsilon_{MoS}$ is plotted as a function of thickness $d_{MoS}$. $\varepsilon_{MoS}$ increases from $3.9\,\varepsilon_0$ for a monolayer to $10.5\,\varepsilon_0$ for bulk MoS₂. The errors originate from the measurements of sample sizes, thicknesses of MoS₂ and BN, and capacitances of MoS₂.

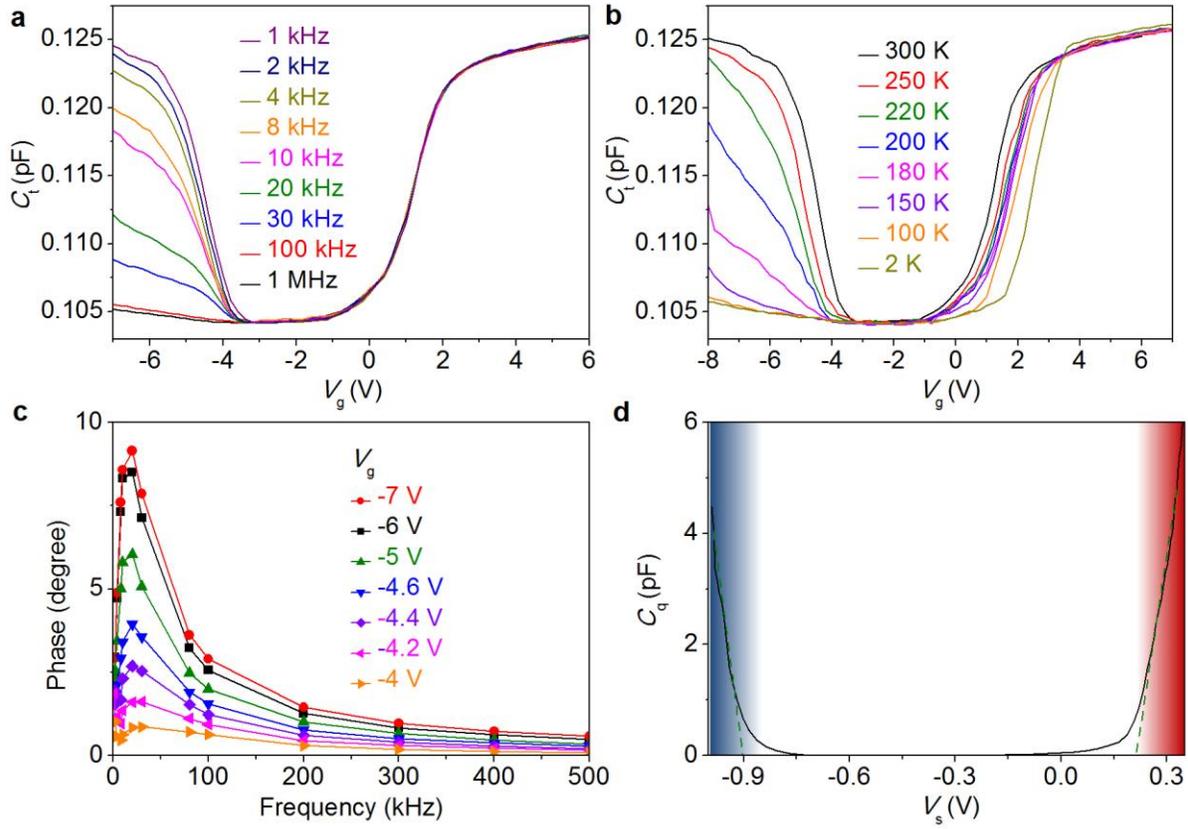

**Figure 3 | The valence band of multilayer MoS$_2$ accessed by capacitance measurement. a**, $C_t$ measured at 300 K for different excitation frequencies. **b**, $C_t$ measured at 1 kHz for different temperatures. **c**, Phase information plotted as a function of excitation frequencies at 300 K for different $V_g$. The phase peak around 20 kHz yields a relaxation time of hole carriers at 50 μS. **d**, The quantum capacitance $C_q$ of MoS$_2$ plotted as a function of surface potential $V_s$ at 300 K, which yields a band gap width of around 1.14 eV. The excitation voltage used is 100 mV.

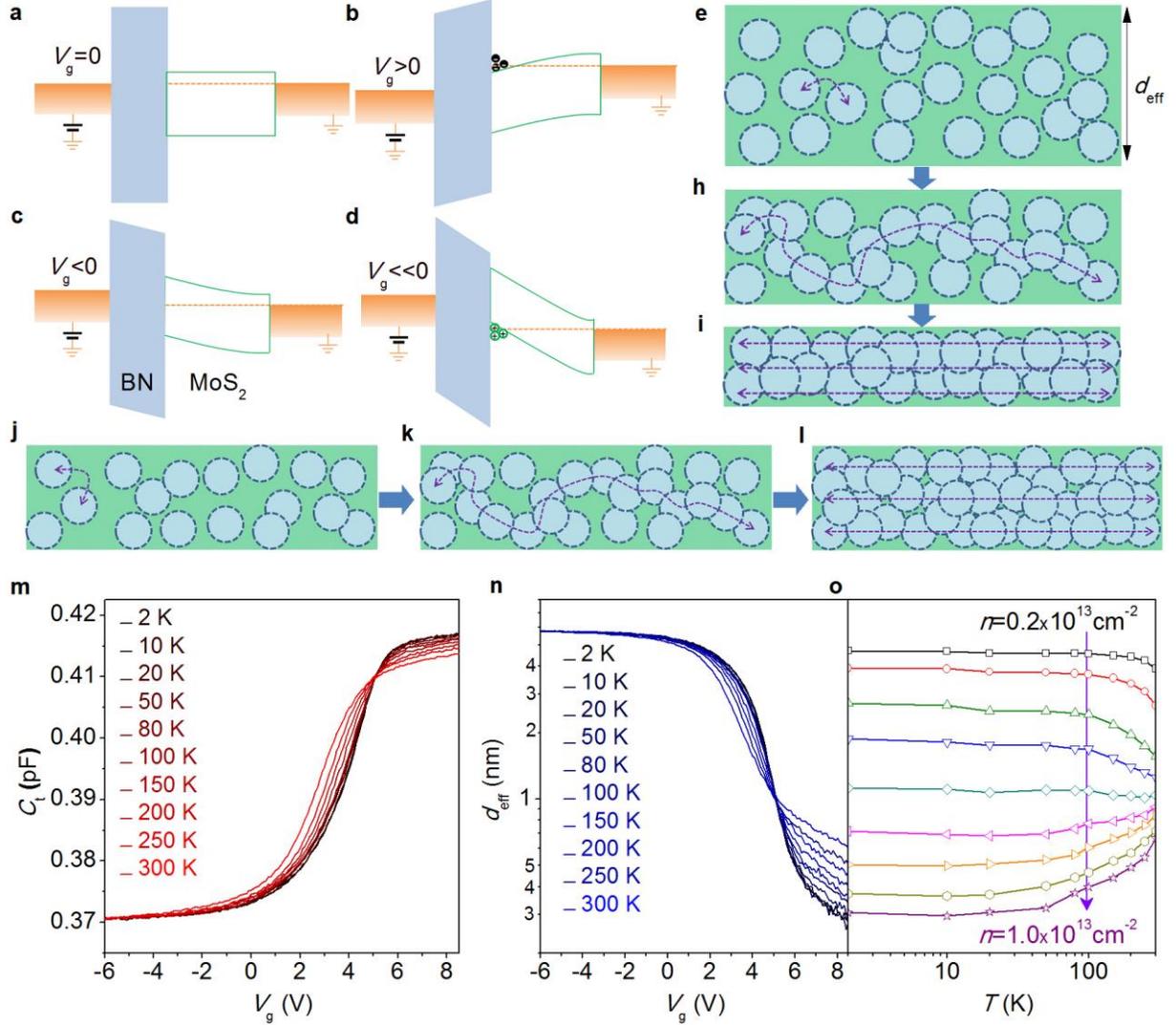

**Figure 4 | The percolation transition driven by density inhomogeneity in multilayer $MoS_2$.** **a-d**, The schematic band diagrams of metal-BN-$MoS_2$-metal structures at flat band (**a**), accumulation region (**b**), depletion region (**c**) and inversion region (**d**). **e-l**, The schematic images showing the percolation-induced MIT under different effective thicknesses of electron states (**e-i**) and carrier densities (**j-l**). The circles denote isolated carrier puddles in $MoS_2$. **m,n**, The measured total capacitance $C_t$ (**m**) and effective thickness $d_{eff}$ (**n**) plotted as a function of gate voltage $V_g$ for 2 K to 300 K. The excitation voltage and frequency used are 50 mV and 100 kHz respectively. **o**, $d_{eff}$ plotted as a function of temperatures at different carrier densities $n$.

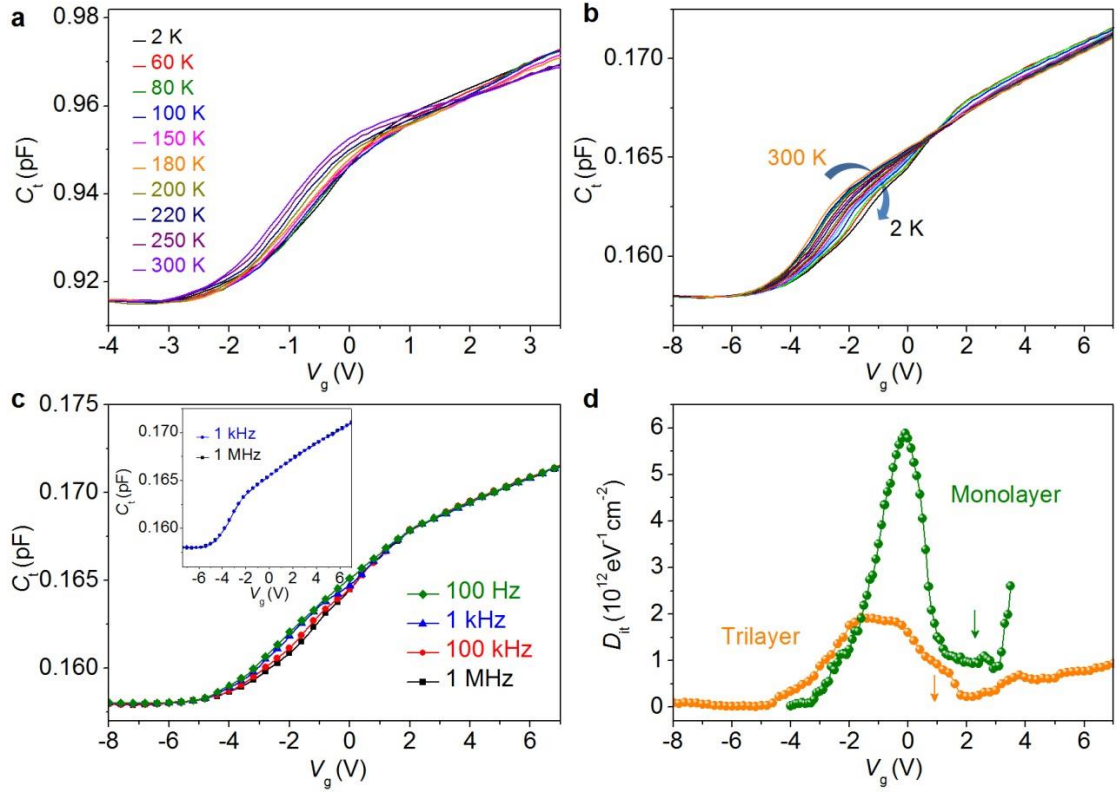

**Figure 5 | The percolation transition and charge traps in monolayer and trilayer MoS$_2$. a**, $C_t$ of a monolayer MoS$_2$ measured at an excitation frequency 1 kHz and excitation voltage 50 mV for different temperatures. **b**, $C_t$ of a trilayer MoS$_2$ measured at an excitation frequency 100 kHz. **c**, $C_t$ of a trilayer MoS$_2$ measured for different excitation frequencies at 2 K, indicating that the charge traps are excited at low frequencies. The inset shows $C_t$ measured at 300 K. **d**, The charge trap densities $D_{it}$ as a function of $V_g$ calculated for the monolayer and trilayer MoS$_2$ samples. The arrows denote the transition points.

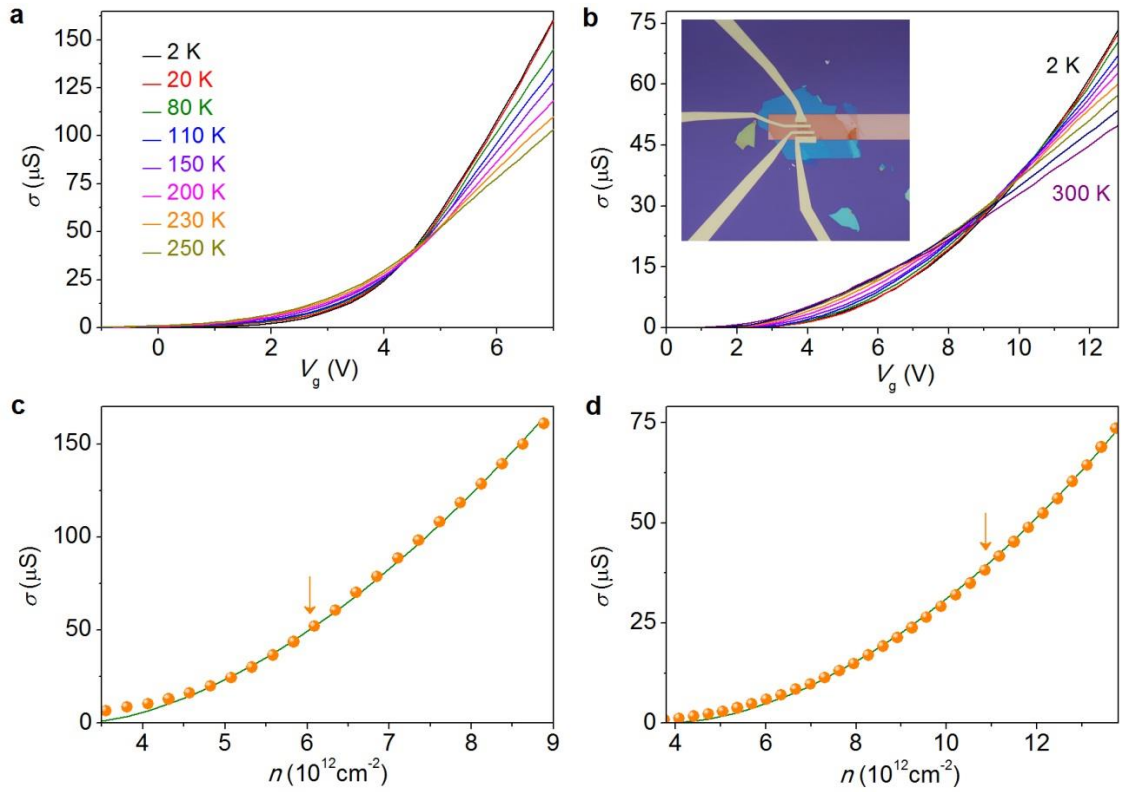

**Figure 6 | Transport results showing the percolation transition in multilayer and monolayer MoS$_2$. a,b**, The MITs are clearly shown by $\sigma$ measurements of a multilayer (**a**) and monolayer (**b**) MoS$_2$ for different temperatures. The inset in (**b**) shows the optical image of a monolayer MoS$_2$ device. **c,d**, The fitting of experimental $\sigma$ (orange dots) of multilayer (**c**) and monolayer (**d**) MoS$_2$ according to the percolation conductivity $\sigma=A(n-n_c)^\delta$ (green lines). The arrows denote the positions of MITs.

# Supplementary Information

## for

## Probing the Electron States and Metal-Insulator Transition Mechanisms in Atomically Thin $MoS_2$ Based on Vertical Heterostructures

Xiaolong Chen[1], Zefei Wu[1], Shuigang Xu[1], Lin Wang[2], Rui Huang[1,3], Yu Han[1], Weiguang Ye[1], Wei Xiong[1], Tianyi Han[1], Gen Long[1], Yang Wang[1], Yuheng He[1], Yuan Cai[1], Ping Sheng[1], Ning Wang[1]*

[1]*Department of Physics and the William Mong Institute of Nano Science and Technology, the Hong Kong University of Science and Technology, Clear Water Bay, Hong Kong, China*

[2]*Department of Condensed Matter Physics, Group of Applied Physics, University of Geneva, 24 Quai Ernest Ansermet, CH1211 Geneva, Switzerland*

[3]*Department of Physics and Electronic Engineering, Hanshan Normal University, Chaozhou, Guangdong, 521041, China*

*Correspondence should be addressed to: Ning Wang, phwang@ust.hk

**Capacitance measurements on MoS$_2$-based FET structures**

Based on standard field-effect transistor (FET) structures, the top electrode is partially covered on MoS$_2$ as shown in Fig. S1a-c (a monolayer MoS$_2$ sample). $S_{max}$ represents the overlapping area between the bottom gate and MoS$_2$, and $S_{min}$ represents the overlapping area between the top electrode and bottom gate. In the partially-covered top electrode structure, $S_{min} < S_{max}$. Fig. S1d-g show the capacitance results from a monolayer MoS$_2$ sample partially covered by a Ti/Au electrode. Under negative bias, MoS$_2$ is in the depleted state, and the measured capacitance can be described by $C_{min} = S_{min} \cdot (\frac{d_{BN}}{\varepsilon_{BN}} + \frac{d_{MoS}}{\varepsilon_{MoS}})^{-1}$. With increasing the gate voltage, electrons start to accumulate at the surface of MoS$_2$ and the channel becomes conductive. Then, the measured capacitance approaches $C_{max} = S_{max} \cdot \frac{\varepsilon_{BN}}{d_{BN}}$. The ratio of $C_{max}$ and $C_{min}$ for this sample is $C_{max}/C_{min} \approx S_{max}/S_{min} = 2.2$. This is because the term $\frac{\varepsilon_{MoS}}{d_{MoS}}$ only contributes about one twentieth of the capacitance $C_{min}$, which is negligible in comparison to the capacitance increase caused by the increase of conducting area (from $S_{min}$ to $S_{max}$) of MoS$_2$.

The capacitance measured from the partially-covered MoS$_2$ devices largely depends on temperatures and frequencies as shown in Fig. S1d-g. The capacitance increases with decreasing excitation frequencies. This is consistent with previously reported results measured at 300 K[1]. The frequency-dependent behavior of the partially-covered devices becomes serious at low temperatures (Fig. S1f). At low temperatures and higher frequencies, MoS$_2$ sheets are normally not fully charged. This is attributed to the charge trapping effect or the huge lateral resistance of MoS$_2$ near the band edge. Hence, some intrinsic characteristics of MoS$_2$ are smeared, especially at low temperatures.

**Extracting the parallel capacitance $C_p$**

To accurately determine the charge trap densities, dielectric constant and quantum capacitance of MoS$_2$, the parallel capacitance $C_p$ shown in Fig. 1c has to be determined first. $C_p$ contains two terms: the residual capacitance $C_r$ originating from the measurement setup (the measurement of $C_r$ will be introduced in next section which is negligible compared to $C_{ex}$) and the parallel capacitance $C_{ex}$ originating from the extra area of the top electrode, as shown in Fig. S2.

$S_t$ is the total effective capacitance area enclosed by the purple dash-dot line. $S_{MoS}$ is the effective area of MoS$_2$ enclosed by the red dashed line. The extra area of top electrode is $S_{ex} = S_t - S_{MoS}$. The actual capacitance $C_t' = C_t + C_{ex} + C_r$, where $C_t$ is the capacitance shown in the main text which excludes the extra capacitance $C_{ex} = S_{ex} \cdot \frac{\varepsilon_{BN}}{d_{BN}}$ and $C_r$. Then capacitance $C_t$ equals to $C_t = (C_{MoS}^{-1} + C_{BN}^{-1})^{-1}$ which corresponds to area $S_{MoS}$.

To determine $C_{ex}$, we only need to know the capacitance of BN per unit area, which can be simply obtained through measuring the reference capacitor shown in Fig. 1a in the main text. We have measured the thicknesses of MoS$_2$ and BN sheets by atomic force microscopy (AFM) in order to determine the dielectric constant of MoS$_2$ (Fig. S3). The thickness of BN is 14 nm for 5.9 nm-MoS$_2$ device, 6.0 nm for monolayer-MoS$_2$ device, 14.9 nm for trilayer-MoS$_2$ device, and 13.5 nm for 12 nm-MoS$_2$ device as shown in the main text. The extracted dielectric constant of BN is around 3.1, consistent with previous results[2].

**Determining the residual capacitance $C_r$**

The residual capacitance $C_r$ is accurately determined by a simple method shown in Fig. S4. In the device structure without any overlap between the top electrode and bottom gate (Fig. S4a), there is a non-conductive channel in MoS$_2$ which cannot be tuned by the bottom gate. Charges from electrodes are blocked from reaching the MoS$_2$ area above the bottom gate. Then, the measured capacitance should be the residual capacitance $C_r$ induced by the measurement setup. The optical image of the 5.9 nm-MoS$_2$ sample before covering the top electrode is shown in Fig. S4b. The measured capacitance along with a 3.3 nm-MoS$_2$ sample is plotted in Fig. S4c. We find that the capacitance values of $C_r$ are in the order of ~fF, which is about 3-order smaller than the MoS$_2$ capacitances. Hence, the residual capacitance $C_r$ is ignored in our calculations.

**Capacitance measurements on the MoS$_2$ device with Cr/Au top electrodes.**

The work function of Cr (~4.5 eV) is larger than that of Ti (~4.3eV). Hence, Cr is much closer to the valence band of MoS$_2$, and a smaller relaxation time of holes should be observed. In the MoS$_2$ sample coated with a Cr/Au electrode (Fig. S5a), the formation of the inversion layer is observed at 2 K where the contribution of thermal effects can be neglected. As confirmed by the capacitance results measured at different excitation voltages (Fig. S5b), leakage current (to the electrode contacts) plays an important role. Large excitation voltages can contribute significantly to the formation of holes. As the work function of Cr is about 0.2 eV higher than the electron affinity of MoS$_2$ (~4.3 eV), the relaxation of electrons is also subjected to the Schottky barrier (Fig. S5a,b) which shows frequency- and temperature-dependent behavior in capacitance measurements. At 300 K, the relaxation time of holes is much shorter (~5 µs) than that at low temperatures, while the capacitance measured at the electron side is almost frequency-independent.

**Modeling capacitance devices and extracting the quantum capacitance $C_q$ of thin MoS$_2$ flakes**

Here, we applied the Poisson equation to investigate our capacitance devices quasi-quantitatively and consider a simplified one-dimensional model[3] as shown in Fig. S6a and b:

$$\frac{d^2V}{dx^2} = -\frac{e(n_D^+ - n - p_A^- + p)}{\varepsilon_{MoS}} \tag{1}$$

where $V$ is the potential of MoS$_2$ at position $x$. $n_D^+$ and $p_A^-$ are the densities of electron and hole donors respectively. $n$ and $p$ represent the electron and hole carrier densities. Assuming $n_0$ and $p_0$ are the equilibrium densities of electrons and holes respectively (at $V = 0$), we have $n = n_0 \exp(\frac{eV}{kT})$ and $p = p_0 \exp(-\frac{eV}{kT})$, where $k$ is the Boltzmann constant. When $V = 0$, the charge neutrality condition applies, and we have $n_D^+ - p_A^- = n_0 - p_0$. Here, we only focus on the depletion and accumulation regions of MoS$_2$ devices where $n_0 \gg p_0$. Then the Poisson equation can be simplified as:

$$\frac{d^2V}{dx^2} = \frac{en_0}{\varepsilon_{MoS}}[\exp(\frac{eV}{kT}) - 1] \tag{2}$$

As the electric field in MoS$_2$ is described by $E = -\frac{dV}{dx}$, the integral of Eq. (2) yields:

$$E = \sqrt{E_0^2 + \frac{2k^2T^2}{e^2L_D^2}[\exp(\frac{eV}{kT}) - \frac{eV}{kT} - 1]} \tag{3}$$

where $L_D = \sqrt{\dfrac{\varepsilon_{MoS} kT}{e^2 n_0}}$ is the Debye length. $E_0$ is the electric field at position $x = d_{MoS}$ (the interface between MoS$_2$ and top electrode shown in Fig. S6a,b). $E_0$ is a non-zero value and mainly determined by the gate voltage $V_g$ because of the finite thickness of MoS$_2$. $E_0$ can be obtained through the constraint condition $\int_0^{V_s} \dfrac{dV}{E} = d_{MoS}$, where $V_s$ is the surface potential of MoS$_2$. As a result, any additional charges $Q_s' = \varepsilon_{MoS} E_0$ will be accumulated at the interface between the top electrode and MoS$_2$ (Fig. S6a,b). The total charges induced in the device can be described by $Q_t = \varepsilon_{MoS} E_s = Q_{MoS} + Q_s'$, where $E_s$ is the surface electric field of MoS$_2$, and $Q_{MoS}$ is the surface charges in MoS$_2$.

The simulating parameters for Fig. S6, S7, and S8 are: $\varepsilon_{MoS} = 8.6$, $d_{MoS} = 5.9\,\text{nm}$, $\varepsilon_{BN} = 3$, $d_{BN} = 13.5\,\text{nm}$ and doping density $n_D^+ = 10^{17}\,\text{cm}^{-2}$. The simulated total capacitance $C_t = \dfrac{dQ_t}{dV_g}$ as a function of $V_g$ is shown in Fig. S6c. Compared to $C_t$ of MoS$_2$ with an infinite thickness, $C_t$ with a finite value of $d_{MoS}$ approaches a constant value $\left(\dfrac{d_{MoS}}{\varepsilon_{MoS}} + \dfrac{d_{BN}}{\varepsilon_{BN}}\right)^{-1}$ in the depletion region, consistent with the experimental observations shown in the main text. The simulations for charge and potential distributions in MoS$_2$ are shown in Fig. S7. In the accumulation region, most of charges are distributed in the first several layers of MoS$_2$. While in the depletion region, the charge density is small (mainly originated from the density of electron donors $n_D^+$). As a result, the charge distribution is almost constant in MoS$_2$. This analysis also supports that the effective thickness $d_{eff}$ decreases with increasing the gate voltage.

The capacitance of MoS$_2$ $C_{MoS} = \dfrac{dQ_t}{dV_s}$ can be written as $C_{MoS} = C_q + C_s{}'$, where $C_q = \dfrac{dQ_{MoS}}{dV_s}$ is the quantum capacitance of MoS$_2$ and $C_s{}' = \dfrac{dQ_s{}'}{dV_s}$ originates from the charges induced at the interface between the top electrode and MoS$_2$. $C_{MoS}$ can be obtained from the serial connection relationship $C_t = (C_{MoS}{}^{-1} + C_{BN}{}^{-1})^{-1}$. However, it is difficult to directly determine $C_s{}'$ as discussed below.

Here, we present an approximate method to satisfactorily extract the quantum capacitance ($C_q$) of MoS$_2$, supported by theoretical calculations. The accurate $C_{MoS}$, $C_s{}'$ and $C_{q\_th}$, without adopting any approximation, are theoretically simulated and shown in Fig. S8. In the depletion region, both $C_s{}'$ and $C_{MoS}$ approaches a constant value $C_{s0}$. With increasing the gate voltage, $C_s{}'$ decreases due to the stronger screening effect of accumulated charges at the surface of MoS$_2$. While, $C_{q\_th}$ increases rapidly. The quantum capacitance of MoS$_2$ can be approximately written in the form of $C_{q\_ex} = C_{MoS} - C_{s0}$ instead of $C_{q\_th} = C_{MoS} - C_s{}'$ with a good accuracy as shown in Fig. S8. In the depletion region, all charges form at the interface between the top electrode and MoS$_2$. Then, we have $C_{MoS} = C_s{}'$, and $C_{q\_th} = C_{q\_ex} = 0$. The approximation is accurate in the depletion region. In the accumulation region, $C_s{}' < C_{s0}$ and there exists deviation between the approximation value $C_{q\_ex}$ and the accurate value $C_{q\_th}$. In this case, surface charges dominate the total induced charges in the device and thus $C_{q\_th} \gg C_s{}'$. Obviously, the deviation is small compared to the value of $C_{q\_th}$. Overall, $C_{q\_ex} = C_{MoS} - C_{s0}$ is a good approximation to the accurate value of $C_{q\_th}$.

**Figures**

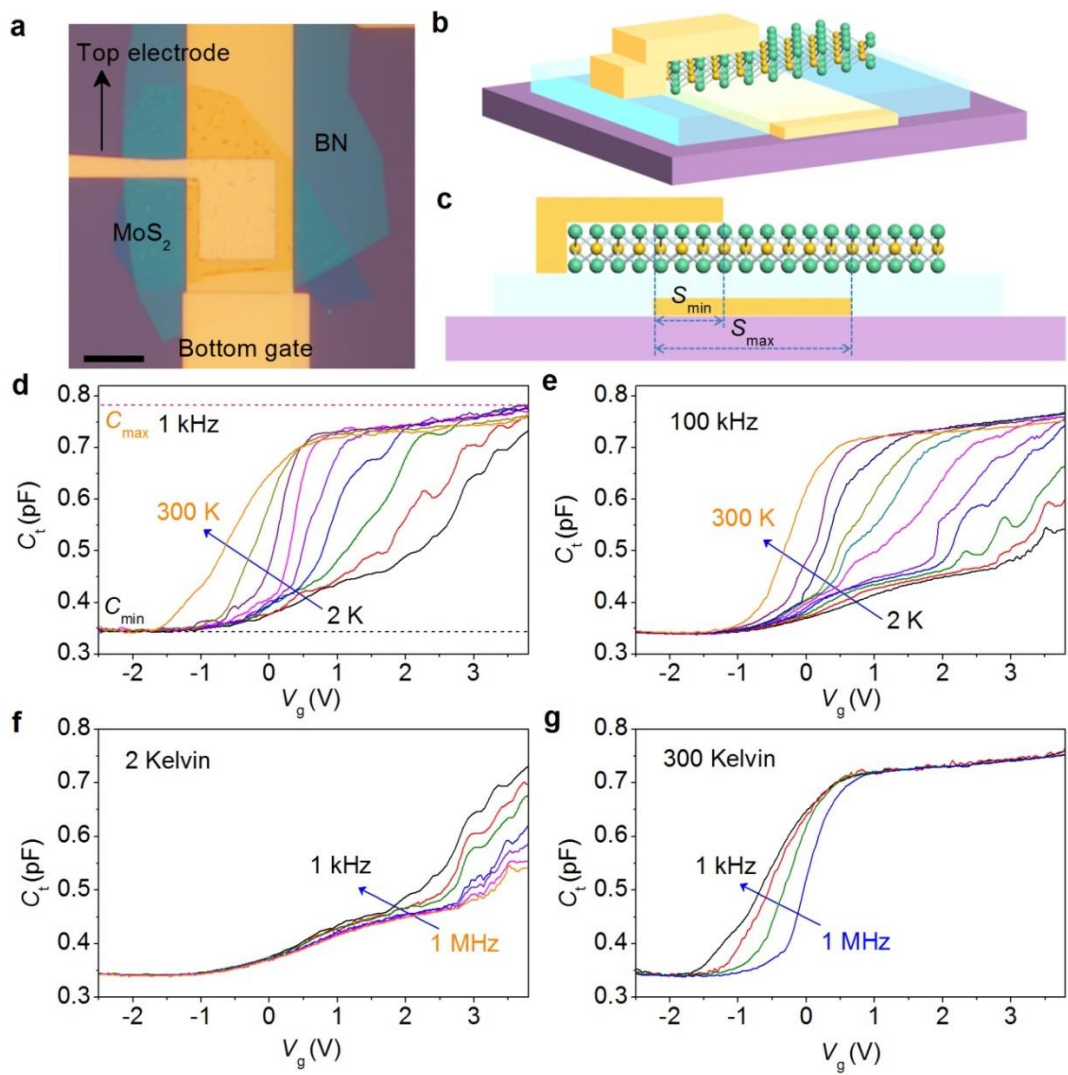

**Fig. S1 | Capacitance characteristics for a monolayer MoS$_2$ in MIS-FET geometry. a-c**, Optical (**a**) and schematic images (**b**,**c**) of the partially-covered top electrode geometry for capacitance measurements. Scale bar is 10 μm. **d**,**e**, $C_t$ at excitation frequency 1 kHz (**d**) and 100 kHz (**e**) for different temperatures. **f**,**g**, $C_t$ at temperature 2 K (**f**) and 300 K (**g**) for different excitation frequencies.

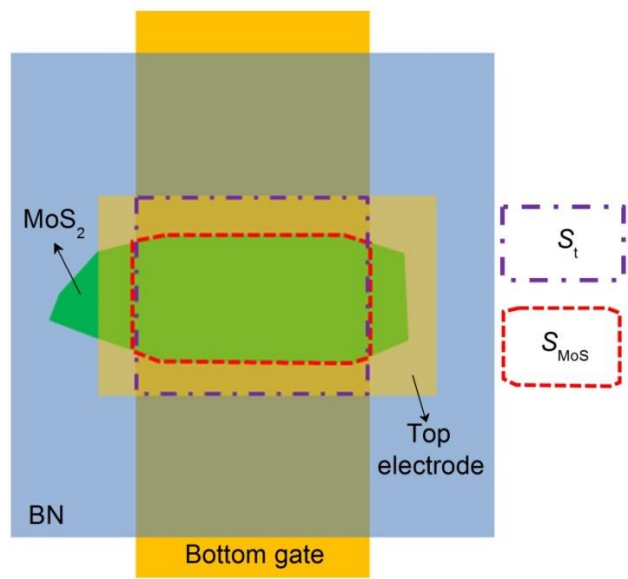

**Fig. S2 | Analysis of the device geometry.** $S_t$ is the total effective capacitance area enclosed by the purple dash-dot line. $S_{MoS}$ is the effective area of MoS$_2$ enclosed by the red dashed line.

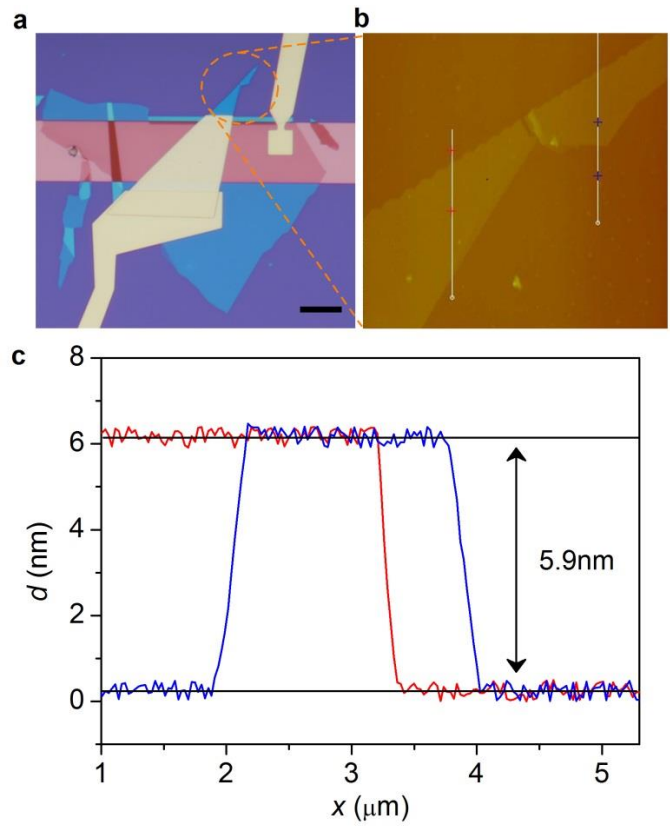

**Fig. S3 | Determination of the thickness of MoS$_2$ by AFM. a**, Optical image of the 5.9nm-thick MoS$_2$ sample shown in the main text. **b**,**c**, AFM image (**b**) and the thickness (**c**) of the MoS$_2$ sample.

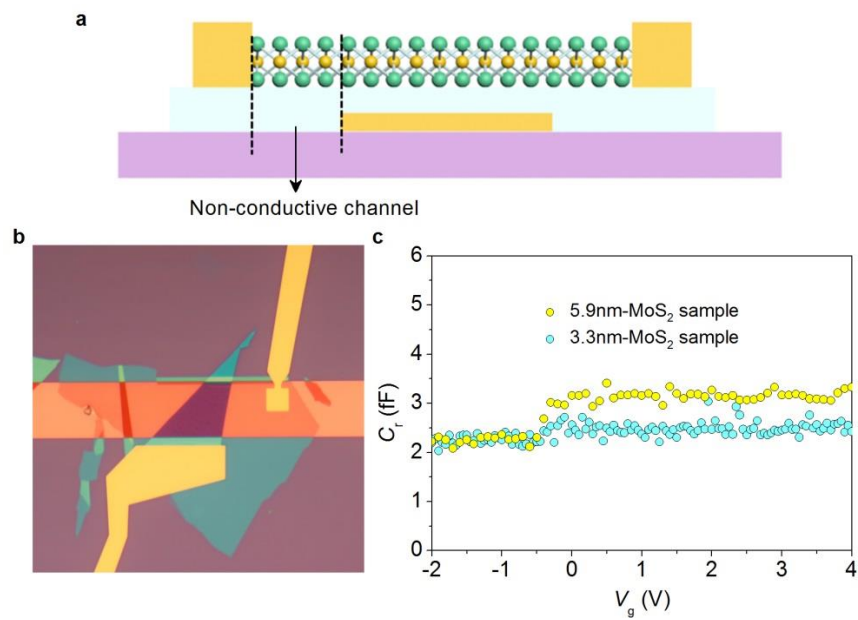

**Fig. S4 | Determination of the residual capacitance $C_r$. a**, Schematic image of the MoS$_2$ sample without overlapping between the top electrode and bottom gate. **b**, Optical image of the 5.9 nm-MoS$_2$ sample before covering the top electrode. **c**, Residual capacitance determined in the 5.9 nm-MoS$_2$ sample along with a 3.3 nm-MoS$_2$ sample. It is three-order smaller than the capacitance of MoS$_2$ devices.

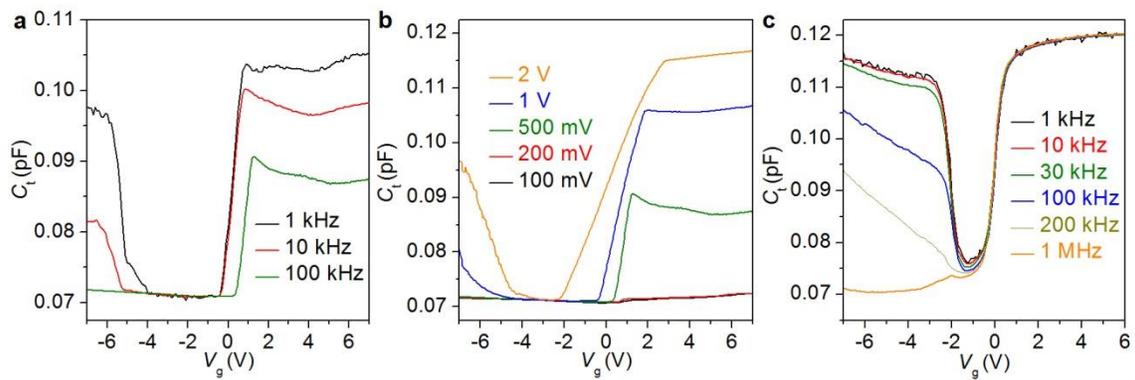

**Fig. S5 | MoS$_2$ capacitance with Cr/Au electrodes. a**, Measured capacitance $C_t$ at 2 K with an excitation voltage 500 mV for different frequencies. **b**, $C_t$ at 2K with an excitation frequency 100 kHz at different excitation voltages. **c**, $C_t$ at 300 K with an excitation voltage 200 mV for different frequencies.

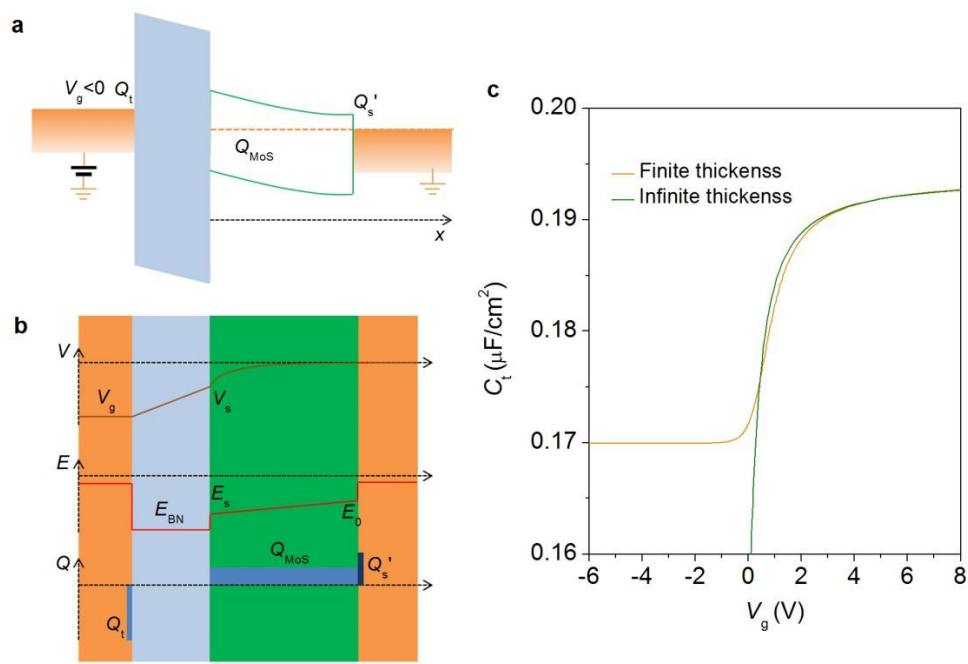

**Fig. S6 | Modeling of capacitance devices. a**, Schematic band diagram at the depletion region. **b**, Schematic image of potentials, electric fields, and charge distributions in the capacitance device. **c**, Simulated total capacitance $C_t$ as a function of gate voltage $V_g$ with a finite thickness $d_{eff} = 5.9\text{nm}$ (orange line) and an infinite thickness (green line).

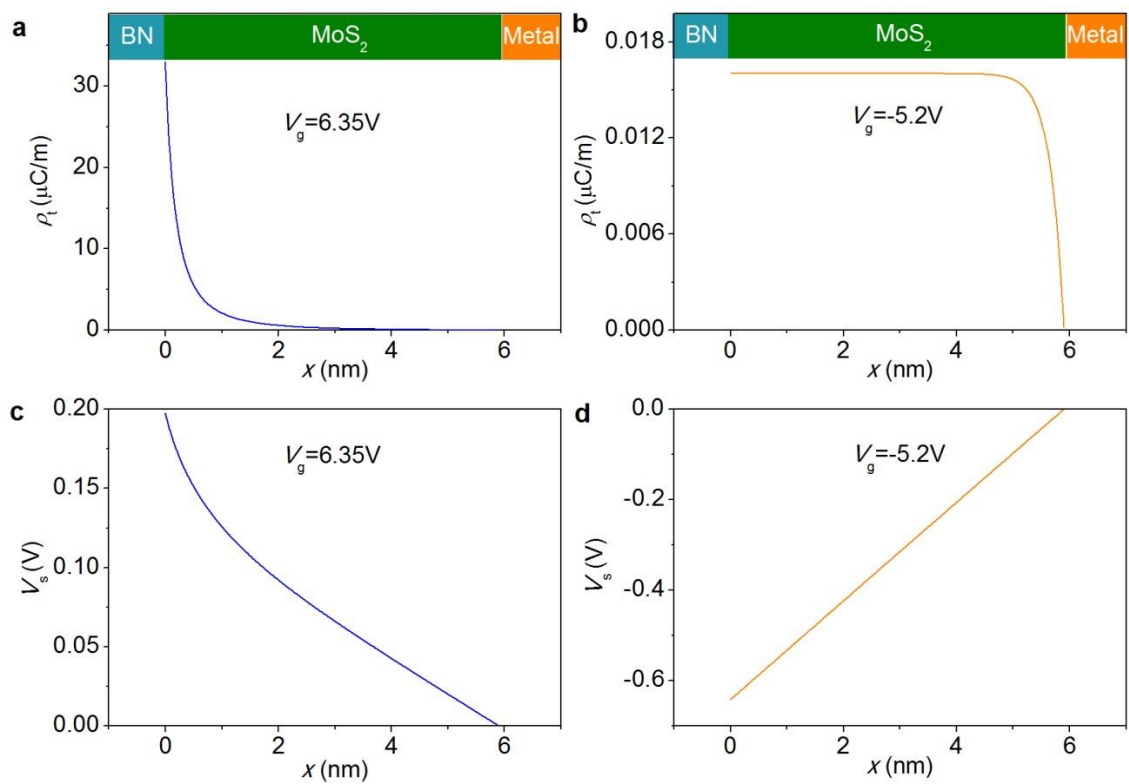

**Fig. S7 | Charge and potential distributions in MoS$_2$. a**,**c**, Charge (**a**) and potential (**c**) distributions in the accumulation region with a gate voltage $V_g = 6.35\text{V}$ in the MoS$_2$ capacitance device. **b**,**d**, Charge (**b**) and potential (**d**) distributions in the depletion region with a gate voltage $V_g = -5.2\text{V}$.

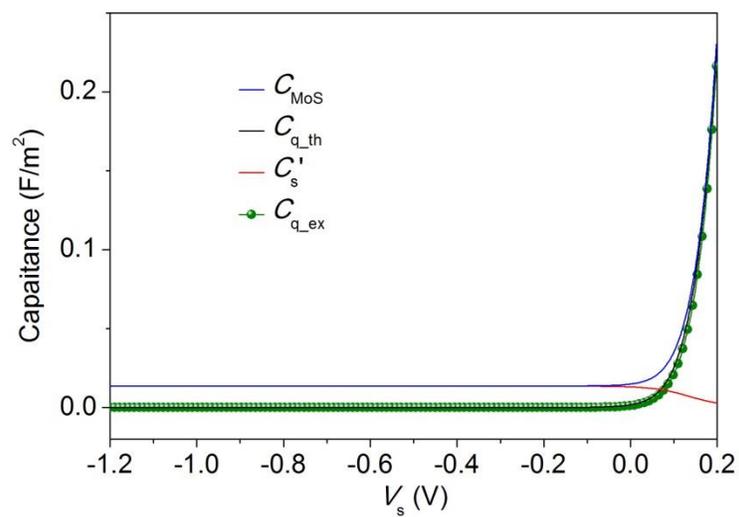

**Fig. S8 | Extraction of the quantum capacitance of MoS$_2$.** Simulated results of $C_{MoS}$, $C_s{'}$ and $C_{q\_th}$ without adopting any approximation. $C_{q\_ex}$ is a good approximation of $C_{q\_th}$.